# Valley-Hall-like second-order photonic topological insulators in Kagome lattice

Le Zhang

*Key Laboratory of Electromagnetic Wave Information Technology and Metrology of Zhejiang Province, College of Information Engineering, China Jiliang University, Hangzhou 310018, China*

Valley-Hall-like second-order photonic topological insulators are designed in Kagome-lattice photonic crystals with $C_{3v}$ point-group symmetry. The photonic crystal consists of circular air holes in pure dielectric materials. Different from conventional valley-Hall photonic topological insulators characterized by valley Chern numbers, the proposed insulators have topological invariants described by quantized electric polarization. Topological transition can be realized by tuning the structural size and topological edge states appear at the interface between photonic crystals with different topological phases, preserving important features of valley-Hall photonic insulators such as valley transport with little backscattering. The proposed photonic crystal also support zero-dimensional corner states in oblique corners, showing its second-order topological insulator signature. This work presents the possibility to realize topologically protected reflection suppressed waveguides and local cavities in the same platform.

## I. INTRODUCTION

Topological photonics provide attractive approaches to reduce photon losses in waveguides or cavities thus gain great developments in recent years. Photonic topological insulators (PTIs), as the extension of topological insulators from electronics to optics, have been realized using different mechanism such as photonic quantum anomalous Hall (QAH) [1-8], quantum spin Hall (QSH) [9-16] or quantum valley Hall (QVH) effects [17-22]. Due to small magnetic response of magneto-optical materials in optical band, great efforts have been made in exploring PTIs which do not need external magnetic field to break the time reversal symmetry. A powerful solution is discovered by applying the concept of valley degree of freedom from valleytronics [23-25] to photonic crystals (PCs). By introducing structural asymmetry in the photonic lattice, the Berry curvature: the "magnetic field" in momentum space, can be nonvanishing and opposite signed at the time-reversal valleys ($K$ and $K'$), resulting in nonzero valley Chern number exhibiting nontrivial topological phase [17-19]. Because of zero coupling between forward-moving $K$-valley and backward-moving $K'$-valley fields, robust topological waveguiding with reflection suppressed through sharp turns is possible, which is newly experimentally demonstrated at telecommunication band [22].

In recent publications, PTIs in the absence of Berry curvatures are reported in a simple square lattice [26,27]. Their topological invariants are characterized by two-dimensional (2D) electric polarization related to 2D Zak's phase [28,29]. Although this kind of PTIs can not avoid backscattering at turns, they reveal a non-vanishing Berry curvature may not be necessary for topologically nontrivial systems. Supplementary understandings of the classical QSH PTIs in hexagonal-lattice PCs with $C_6$ point-group symmetry (PGS) [13] are explained using this theory in their earlier reports [30]. Those square-lattice PTIs with zero Berry curvature, which can be considered as 2D photonic generalization of the Su-Schrieffer-Heeger (SSH) model, are also used to construct higher-order topological insulators (HOTIs) [31]. In the second-order PTIs, besides one-dimensional edge states there are important zero-dimensional corner states. During the writing process of this paper, observation of higher-order topological acoustic states in Kagome-lattice acoustic metamaterial is reported [32,33].

In this work, we propose a valley-Hall-like photonic topological insulator that owns many similar properties with conventional valley-Hall PTIs. The designed PTI is made of periodic circular air-hole array in pure dielectrics, forming a Kagome-lattice photonic crystal with $C_{3v}$ point-group symmetry. We theoretically demonstrated its topology can transit by shrunk or expand the $C_{3v}$ PGS unit cell. The topological nontrivial phase of its photonic band gap is quantized by a fractional electric polarization $(\frac{1}{3}, \frac{1}{3})$ connected to 2D Zak's phase. Numerical simulations illustrate topological 1-D edge states can emerge at the interface between two PCs with different topological phases. Unlike those square-lattice HOTIs with non-negligible reflection at turns, the proposed PTI preserves the key features of valley transport with little reflection at turns along the same valley. Furthermore, 0-D corner states can exist in an oblique corner of the proposed PTI, showing higher-order PTI properties. The principle can also be used to construct rod-type PTIs with similar features in other electromagnetic bands through proper structural designs.

## II. QUICK REVIEW ON PHOTONIC VALLEY-HALL PHASE

At the beginning, we present a quick review of a valley hole-type photonic crystal (PC). The unit cell consists of two air holes embedded in pure dielectrics with the hole diameters of $d_1$ and $d_2$ respectively in hexagonal lattice, as shown the hexagon in Fig.1(a). The permittivity of the dielectric is $\varepsilon=13$. We fix the size $d_1=0.4a$ where $a$ is the lattice constant and change the size $d_2$ of the center hole of the unit cell. The photonic transverse-electric (TE) mode band structure of the lowest three bands are shown in Fig.1(b),(c),(d) under three different $d_2$. When $d_2=0.2a<d_1$, there is a band gap opening between the lowest two bands. When $d_2=0.4a=d_1$, the lowest gap closes and reduces to a degeneracy featuring a Dirac cone at $K(K')$ point in the momentum space, as pointed out with a red dash circle in Fig.1(c). When $d_2=0.6a>d_1$, the lowest gap opens again.

The $n$th-band Berry curvature is defined as $F_{\mathbf{k}}^{(n)} = \nabla_{\mathbf{k}} \times A_{\mathbf{k}}^{(n)}$ with $A_{\mathbf{k}}^{(n)}$ denoting the Berry connection of the $n$th band energy $u_{n\mathbf{k}}$ [34]. By using an efficient algorithm [35], we did a fast calculation on the Berry curvature of the first band over a roughly discretized portion of the Brillouin zone (BZ). The asymmetry between the sizes of the two holes ($d_1 \neq d_2$) within the unit cell breaks the inversion symmetry, leading to non-zero Berry curvature with opposite signs at the $K$ and $K'$ points. This causes non-zero valley Chern numbers thus both the cases for $d_2=0.2a$ and $d_2=0.6a$ show topologically nontrivial valley-Hall phase. Above valley-hall physics have already been discussed in many publications related to the valley Hall PTIs [17-22]. It is worth to mention an exchange process of the Berry curvature with respect to the band inversion. If $d_2$ starts increasing from $d_2<d_1$, a band inversion between the 1st and 2nd bands will take place at the band degeneracy when $d_2=d_1$. As compared the fields in Fig.1(e), bands 1, 2 and 3 in $d_2=0.2a$ case correspond to band 3, 1 and 2 in $d_2=0.6a$ case respectively, clearly showing the inversion. As arrowed the power flux, the helicity of the lowest mode is usually used to explain the backward suppression of valley transports because zero coupling is verified between the field distribution of left-hand circular polarization (LCP) and right-hand circular polarization (RCP) modes [17]. As shown in Fig.1(f), the Berry curvature at $K(K')$ point will have an exchange between lowest two bands after the inversion, resulting in a reversion of sign of the Berry curvature near $K(K')$ point. The valley Chern number will then change its sign from $C_{K/K'}=\pm 0.5$ to $C_{K/K'}=\mp 0.5$. In addition, simply rotating the original unit cell by 180° will also reverse $C_{K/K'}$ from $\pm 0.5$ to $\mp 0.5$. Topological edge states can emerge at the edge between two PCs with different valley Chern numbers.

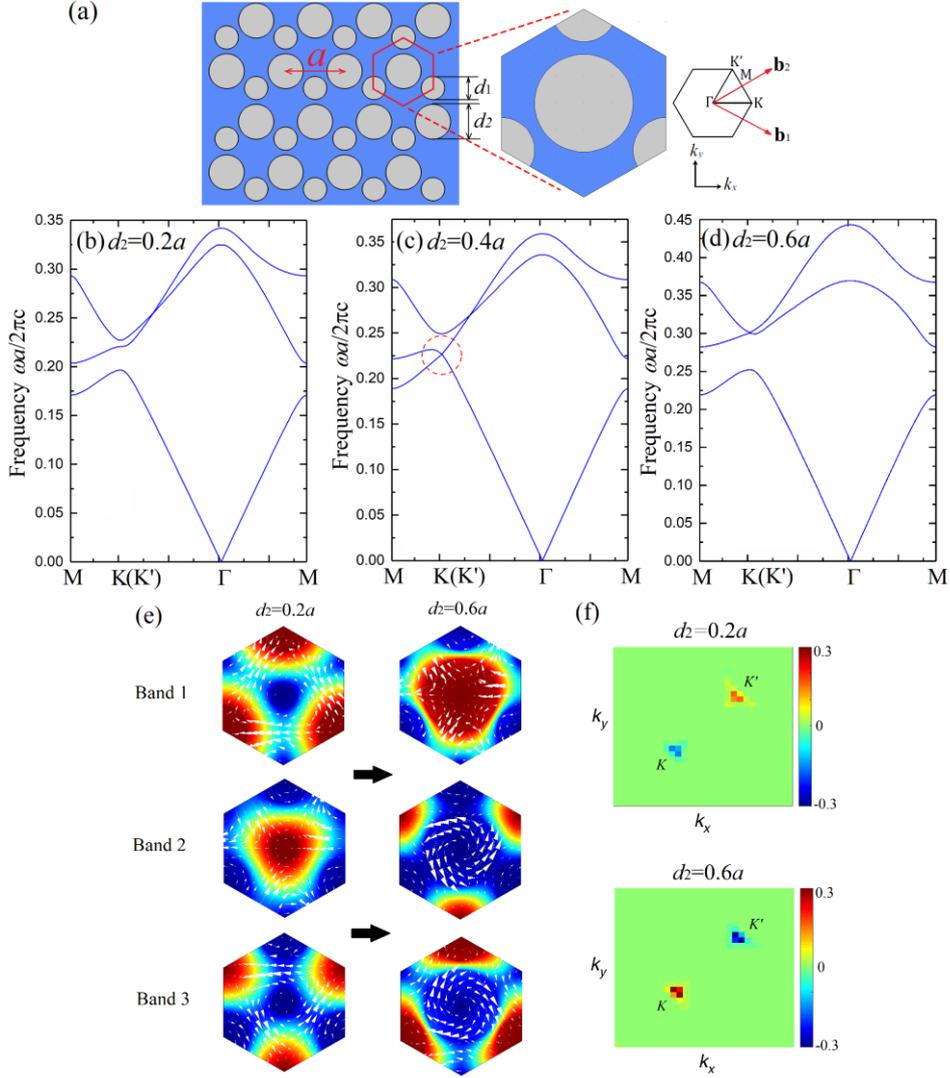

FIG.1. (a) Typical structure of a valley hole-type photonic crystal; (b-d) Band structures of the photonic crystal with the diameter of center air hole in the cell $d_2$=0.2$a$, 0.4$a$, 0.6$a$ respectively; (e) Normalized field distribution of the lowest three bands at K point under $d_2$=0.2$a$ and $d_2$=0.6$a$ cases, showing the band inversion. The white arrows present the direction of power flux; (f) Berry curvature of the first band under $d_2$=0.2$a$ and $d_2$=0.6$a$ cases.

### III. VALLEY-HALL-LIKE PHOTONIC TOPOLOGICAL INSULATOR

Then we consider a Kagome-lattice photonic crystal, the unit cell of which is composed of three identical air holes with $C_{3v}$ PGS embedded in pure dielectrics, as shown in Fig.2(a). The permittivity of the dielectric is $\varepsilon$=13. We set the diameter of the holes $d$=0.36$a$ where $a$ is the lattice constant. We explore the topological phase transition by a method of shrinking or expanding the air-hole array within the unit cell, similar to the method in the publications [13]. According to the 2D Su-Schrieffer-Heeger (SSH) model, the two parameters of the intracellular coupling $\gamma$ and the intercellular coupling $\gamma$' determine the topology of the system. Thus the dominant thing in our system is the same distance $W$ between the center of each air hole and the center of the unit cell. One can easily find $W = a/(2\sqrt{3})$ as a critical value when the intracellular coupling equals to the intercellular coupling ($|\gamma'|/|\gamma|$=1).

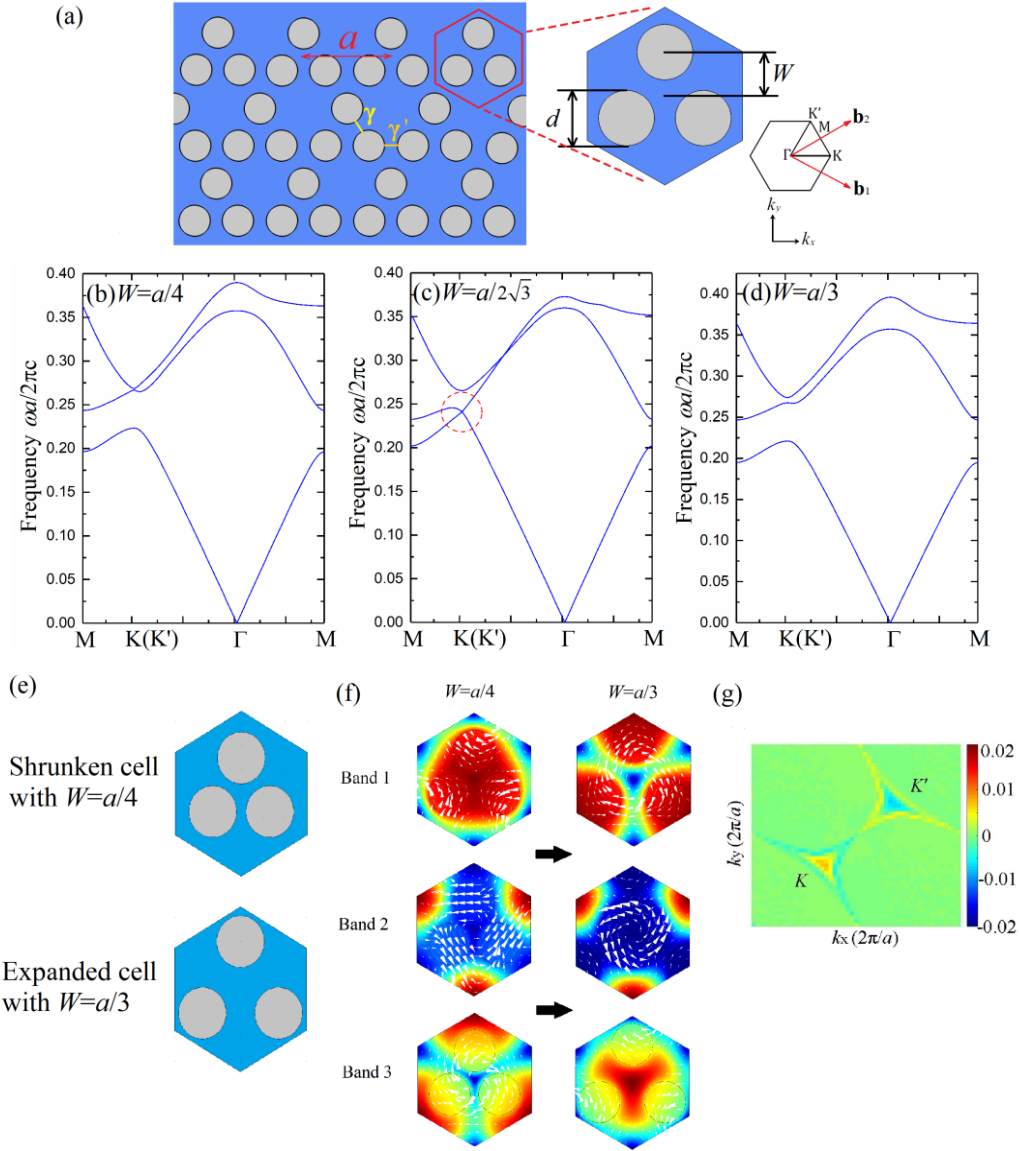

FIG.2. (a) The structure of a hole-type photonic crystal with three identical air holes in the unit cell forming Kagome lattice; (b-d) Band structures of the photonic crystal with the distance between the center of each air hole and the center of the unit cell $W = a/4, a/(2\sqrt{3}), a/3$ respectively; (e) The schematics of a shrunken unit cell with $W=a/4$ and a expanded unit cell with $W=a/3$; (f) Normalized field distribution of the lowest three bands at K point under $W=a/4$ and $W=a/3$ cases, showing the band inversion. The white arrows present the direction of power flux; (g) The Berry curvature of the first band is small near $K(K')$ point in Kagome lattice.

With this specific structural size of the photonic crystal, the photonic band structure of the lowest three bands are shown in Fig.2(c). Here, we only focus on the TE modes propagating in the x–y plane. Obviously there are Dirac cones formed by the lowest two bands as marked with a red circle at $K(K')$ point in the first BZ. We then vary the distance $W$ with respect to a to get the three holes shrunken ($W < a/(2\sqrt{3})$) or expanded ($W > a/(2\sqrt{3})$). The corresponding band structures for $W=a/4$ and $W=a/3$ are shown in Fig.2(b) and (d) respectively. The shrunken and expanded unit cells are shown in Fig.2(e). Comparing Fig.2 with Fig.1, the transformation of the bands has many things in common with that of the valley-hall PTI in Sec.II. During the expanding process of $W$ from $a/4$ to $a/3$, band inversions take place between the 1st and 2nd bands at $K(K')$ degeneracy when $W = a/(2\sqrt{3})$. The variations of the field at $K$ in band 1, 2 and 3 are illustrated in Fig.2(f), showing the inversion. The process makes the original existing lowest band gap start from open to closed and then reopened. Through computation, it is found that the 1st-band Berry curvature near $K(K')$ point in the BZ is very small, which is different with the case of ordinary valley-Hall PTI in Sec.II. Since the band inversions will not cause large exchange of the curvature at $K(K')$ points, the traditional valley Chern number here is not $\pm 0.5$. We further show valley Chern number here is not a topological invariant (see the Supplemental Material [38]), thus the topology of a Kagome system is not suitable to be described by the valley Chern numbers. We demonstrate a 2D electric polarization as a complimentary quantum number to characterized the topological transition in our PCs. The 2D polarization $\mathbf{P}= (p_1, p_2)$ is given by the integral of the Berry connection over the 2D Brillouin zone and expressed as [29]

$$p_i = -\frac{1}{(2\pi)^2} \iint_{BZ} dk_1 dk_2 \, \text{Tr}\left[A_i(\mathbf{k})\right], \tag{1}$$

with $i$ indicating the component of $\mathbf{P}$ along the reciprocal lattice vector $\mathbf{b}_i$ ($i=1,2$). Here $\left[A_i(\mathbf{k})\right]^{mn} = -i\langle u^m(\mathbf{k}) | \partial_{k_i} | u^n(\mathbf{k}) \rangle$ is the Berry connection matrix where $m$ and $n$ run over occupied energy bands. $|u^n(\mathbf{k})\rangle$ is the periodic Bloch function for the $n$th band. $\mathbf{k} = k_1 \mathbf{b}_1 + k_2 \mathbf{b}_2$ is the wave vector. Via a direct integration with Eq.(1), it is found the 2D polarization $\mathbf{P}$ of the proposed PC along $\mathbf{b}_{1,2}$ (as illustrated in Fig.2(a)) equals $(\frac{1}{3},\frac{1}{3})$ for $W > a/(2\sqrt{3})$ showing topological nontrivial phase, while it equals (0,0) for $W < a/(2\sqrt{3})$ showing topological trivial phase. The calculation is conducted using the Wilson-loop approach following Refs.[36,37]. Its mathematical details are presented in the Supplemental Material [38]. This polarization can also be obtained by making use of $C_3$ PGS. Under the specific PGS, the 2D polarization can be written as [29]

$$\exp(i2\pi p_{1,2}) = \prod_{n \in occ} \frac{\beta_n(K)}{\beta_n(K')}, \tag{2}$$

where $\beta_n(\mathbf{k})$ is the eigenvalue of the operator $R_3$ at $C_3$-invariant $\mathbf{k}$ point on the $n$th band. Its derivation is given in the Supplemental material [38].

Then, we investigate the topological edge modes between two PCs with different 2D polarization. Four types of edges supporting topological edge modes are considered with their supercell structure and projected band diagrams along the edge direction (x-axis). The first one in Fig.3(a) is formed by an expanded PC ($W=a/3$) with $\mathbf{P} = (\frac{1}{3},\frac{1}{3})$ and a shrunken PC ($W=a/4$) with $\mathbf{P} = (0,0)$. A topological edge mode proved to be confined at the edge through the normalized field distribution is presented by red line inside the lowest topological photonic band gap. Its time-reversal partners respectively at positive and minus $k_x$ region are in y-axial symmetry but always with opposite group velocity. The edge mode does not connect the upper and lower bulk bands. The second edge in Fig.3(b) is formed by simply rotating the expanded PC of the first edge structure by 60 degree, leading to the topological nontrivial polarization becoming $\mathbf{P} = (-\frac{1}{3},-\frac{1}{3})$. The unique topological edge mode also appears but the field seems less confined at the interface comparing to the first edge. The third edge in Fig.3(c) is constructed by substituting the upper shrunken PC of the first edge structure with ordinary triangular array of air holes (also topologically trivial) with the diameter of 0.6*a*, so the properties of the edge mode have no much difference with the first edge. The last edge in Fig.3(d) consists of two expanded PCs. The lower region is still the expanded PC with $\mathbf{P}=(\frac{1}{3},\frac{1}{3})$, while the upper is the expanded PC with 60°-rotated unit cell possessing $\mathbf{P}=(-\frac{1}{3},-\frac{1}{3})$. Two topological edge modes in the gap is supported by such an edge structure, which can not be constructed in square-lattice high order PTIs.

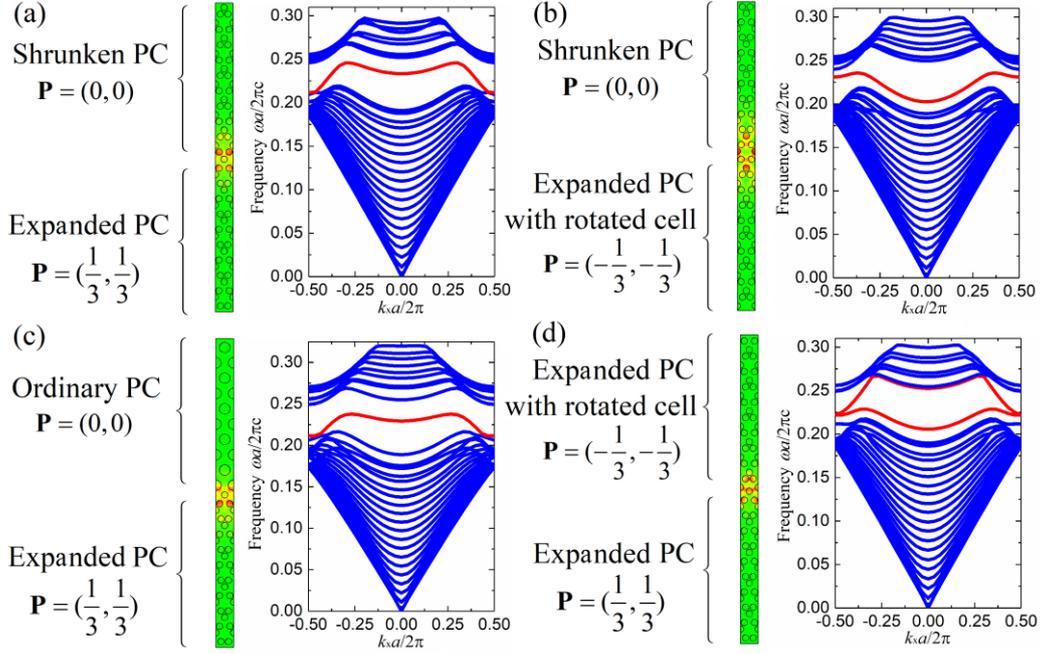

FIG.3. Projected band diagrams and topologically protected states of the edge between two PCs with different 2D polarization. (a) edge between a shrunken PC and an expanded PC; (b) edge between a shrunken PC and an expanded PC with the cell rotating 60°; (c) edge between an ordinary hole-type PC and an expanded PC; (d) edge between an expanded PC and an expanded PC with the cell rotating 60°.

Next, we discuss the wave propagation of the topological edge modes. The time-averaged power flow is computed by using Comsol Multiphysics and the source excites a harmonic wave with the frequency of 0.23(c/*a*) locating inside the band gap. The first situation is a stable straight waveguiding between two expanded PCs with different unit-cell orientations, as shown in Fig.4(a). Such a waveguide structure is quite similar to that of a valley-hall PTI interface between two PCs with different orientations, which is not possible in square-lattice high order PTIs. The second waveguiding between an expanded PC and a shrunken PC through a sharp oblique turn (the same *K* direction) further demonstrates that the proposed PTI preserve key features of the conventional valley-hall PTI. As shown in Fig.4(b), the topological transport is robust through the turn with reflection suppressed, due to zero field coupling between the forward and the backward modes. The mechanism lies on that the lowest band of the expanded nontrivial PC owns helical field properties, as shown the white arrowed power flux at *K* point for *W*=*a*/3 case in Fig.2(f). Obviously the corresponding field properties at *K'* point will show opposite helicity. So ideally there is no field coupling between the forward wave along *K* direction and backward wave along *K'* direction, which is the same as in the valley-Hall PTIs [17]. The corresponding transmission spectrum illustrated in Fig. 4(d) clearly reveals the transmission band with a nearly unitary transmittance larger than 95%. This is also the main reason that we call the proposed PC a valley-hall-like PTI. However in Fig.4(c), when edge waves propagate though a sharp orthogonal turn, larger reflection takes place with the transmittance though the turn reducing to 76%. This is because at orthogonal corner the forward mode can flip to another backward mode with opposite group velocity and helicity. To show the influence brought by disorders, a missing lattice is introduced on the edge between an expanded PC and an ordinary triangular-lattice air-hole PC. As shown in Fig.4(e), the edge waves can go around the defect to some degree but the existing backscattering causes a reduced transmittance of at most 90% through the disorder. Above results demonstrate the valley-like edge transports.

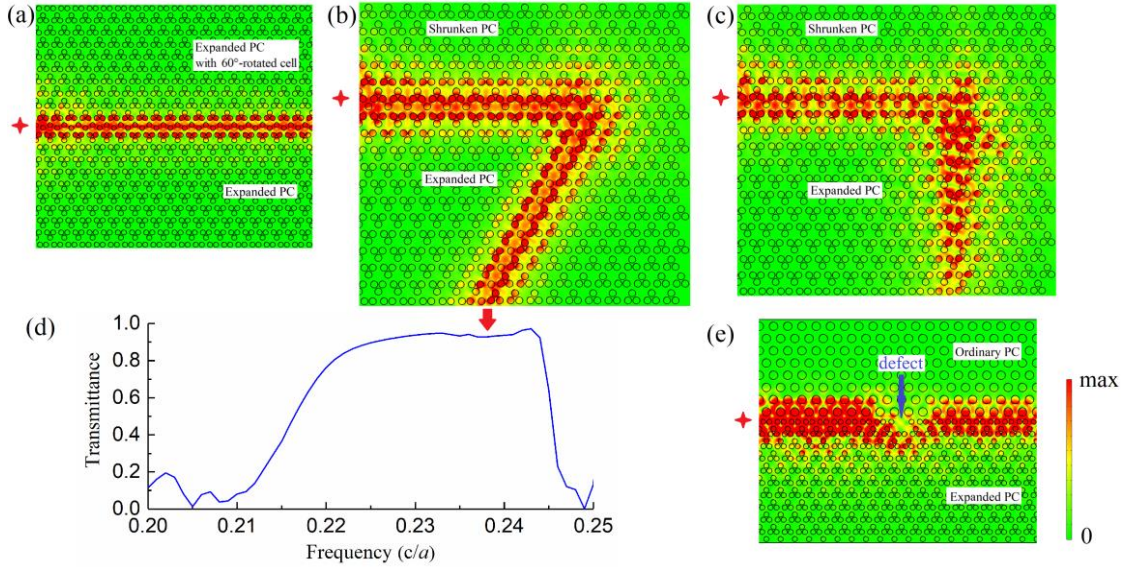

FIG.4. Edge propagation towards different directions. (a) Straight edge guiding between an expanded PC and an expanded PC with the cell rotating 60°. (b) Edge guiding through a sharp oblique turn between a shrunken PC and an expanded PC. (c) Edge guiding through an orthogonal turn between a shrunken PC and an expanded PC. (d) The transmittance over the sharp oblique turn in (b). (e) Straight edge guiding through a disorder formed by removing a unit cell close to the edge.

## IV. CORNER STATES IN SECOND-ORDER PHOTONIC TOPOLOGICAL INSULATOR

The last important thing is, the proposed Kagome-lattice photonic crystals can support zero-dimensional localized corner states in an oblique corner, which is constructed by two edges between topologically non-trivial PC and topologically trivial PC. Through establishing the tight-binding Hamiltonian, to obtain the zero energy states, the eigenvalue equation satisfies $H|\psi\rangle = 0$, where $|\psi\rangle$ is the eigenstate of the zero energy [32]. It is derived that the decay length of the zero energy corner state is

$$l = \log \frac{\gamma'}{\gamma} \qquad (3)$$

with γ and γ' respectively denoting the intracellular and intercellular coupling as mentioned in Fig.2.

If γ' > γ, the decay length *l* is positive thus the corner decay mode is predicted to exist. If γ' < γ the decay length *l* is negative, revealing non-decay modes thus corner modes do not exist. In another word, corner modes can only exist in those corners with decay modes away from the corner along both the edge vector directions. To prove the prediction, we numerically find out the corner state in a 60° corner shown in Fig.6(a) with the expanded PC inside the corner and the shrunken PC outside the corner. The configurations of the expanded and shrunken PC are the same as those in Fig.2. The 0-D corner state locating within the band gap is labeled out with a red dash line in the projected band structure in Fig.5, distinguished from the 1-D edge state in orange solid line. The frequency of the corner state differs from that of the edge state. This corner state is protected by the non-trivial topology, which is a smoking-gun evidence of the second-order photonic topological insulators.

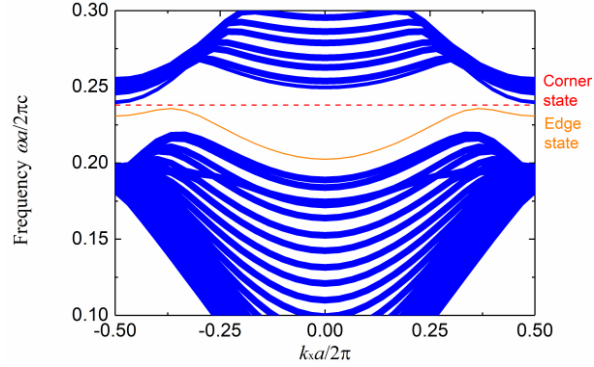

FIG.5. Projected band structure along the $k_x$ direction for a 60° corner like Fig.6(a). The structural size of the PCs is the same as that in Fig.2. The red dashed line labels the frequency of the 0-D corner state, comparing with the 1-D edge state in orange solid line within the band gap.

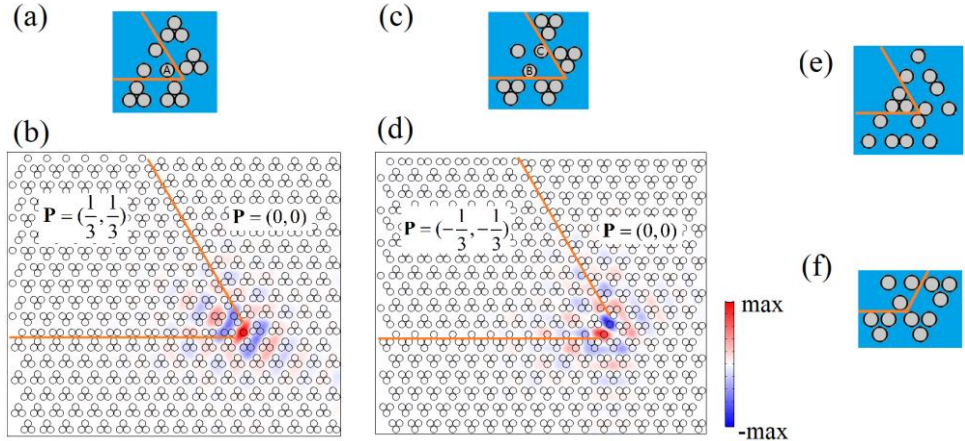

FIG.6. Corner states under different corner constructions. The edges indicating by orange lines are formed between an expanded PC with $W=a/4.6$ and a shrunken PC with $W=a/2.8$. (a) A 60° corner having the expanded PC with $\mathbf{P}=(\frac{1}{3},\frac{1}{3})$ inside the corner and shrunken PC with $\mathbf{P}=(0,0)$ outside the corner. (b) The magnetic fields $H_z$ of the corner state for (a). Fields are strongly localized at the corner near the hole site A. (c) A 60° corner having the expanded PC with $\mathbf{P}=(-\frac{1}{3},-\frac{1}{3})$ inside the corner and shrunken PC with $\mathbf{P}=(0,0)$ outside the corner. (d) The magnetic fields $H_z$ of the corner state for (c). Fields are strongly localized at the corner near the hole site B and C. (e) A 60° corner having the trivial PC inside the corner and non-trivial PC outside the corner. Corner states cannot exist in such a corner. (f) A 120° corner having the non-trivial PC inside the corner and trivial PC outside the corner. Corner states cannot exist in such a corner either.

More detailly, the corner states under four different corner constructions are discussed. It is obvious from Eq.(3) that the corner mode is more localized at the corner with larger γ' / γ ratio. Hence we choose lager γ' / γ ratio to form the corner by setting $W=a/2.8$ of the expanded PC and $W=a/4.6$ of the shrunken PC. The first 60° corner in Fig.6(a) has the expanded PC with $\mathbf{P}=(\frac{1}{3},\frac{1}{3})$ inside the corner and the corner field localized near the hole site A is clearly confirmed by the field distribution simulation in Fig.6(b). The second 60° corner in Fig.6(c) also supports corner states but has the expanded PC with $\mathbf{P}=(-\frac{1}{3},-\frac{1}{3})$ inside the corner. The field distribution in Fig.6(d) shows in such a corner the fields mainly localize near the hole sites B and C. However, the corner states cannot exist in a 60° corner shown in Fig.6(e) having trivial PC inside the corner and non-trivial PC outside the corner because the states grow away from the corner rather than decay. Other corners, such as a 120° corner shown in Fig.6(f) having non-trivial PC inside the corner and trivial PC outside the corner, cannot support corner states either since along one of the lattice vectors the mode is not a decay mode away from the corner. These discussions may be potentially used to guide the implementation of optical cavities or even lasers with highly localized corner modes in oblique corners, which needs further investigations.

## V. Conclusion

We have studied Kagome-lattice photonic crystals with three air holes in the unit cell obeying $C_{3v}$ point-group symmetry. They show valley-hall-like topological phases and their topological properties are protected by 2D electric polarization. They possess many similar characteristics to conventional valley-Hall photonic crystals, such as the edge states in the system here can turn along the same valley through a sharp corner with suppressed backscattering. Such features are unable to exist in square-lattice high order photonic crystal system. The proposed structures of air holes in pure dielectrics are easy to fabricate and can also extend to rod-type photonic crystals with the same point-group symmetry. The last important thing is, hexagonal-lattice higher-order topological insulators with oblique optical corner modes can be realized in the proposed photonic crystals and then may be used to fabricate optical lasers. This work proves the understanding of using fractional electric polarization as a quantum number to describe topology in photonic honeycomb system, and add new sights on achieving valley waveguides and local cavities in the same photonic topological insulators.